# Two-Tier Prediction of Solar Power Generation with Limited Sensing Resource

Yubo Wang[*], Bin Wang[*], Rui Huang[*], Chi-Cheng Chu[*], Hemanshu R. Pota[†] and Rajit Gadh[*]
[*] Department of Mechanical Engineering
UCLA, Los Angeles, USA
[†]School of Engineering and Information Technology
University of New South Wales, Canberra, Australia

*Abstract*—**This paper considers a typical solar installations scenario with limited sensing resources. In the literature, there exist either day-ahead solar generation prediction methods with limited accuracy, or high accuracy short timescale methods that are not suitable for applications requiring longer term prediction. We propose a two-tier (global-tier and local-tier) prediction method to improve accuracy for long term (24 hour) solar generation prediction using only the historical power data. In global-tier, we examine two popular heuristic methods: weighted k-Nearest Neighbors (k-NN) and Neural Network (NN). In local-tier, the global-tier results are adaptively updated using real-time analytical residual analysis. The proposed method is validated using the UCLA Microgrid with 35kW of solar generation capacity. Experimental results show that the proposed two-tier prediction method achieves higher accuracy compared to day-ahead predictions while providing the same prediction length. The difference in the overall prediction performance using either weighted k-NN based or NN based in the global-tier are carefully discussed and reasoned. Case studies with a typical sunny day and a cloudy day are carried out to demonstrate the effectiveness of the proposed two-tier predictions.**

## I. INTRODUCTION

In recent years, regions with rich solar resource such as California and Nevada have committed large scale solar installations at residential, commercial and utility levels. Solar Energy Industry Association (SEIA) estimated that in California, 4,316MW capacity of solar panels were installed in 2014 and the US solar capacity is projected to be over 12,000MW in 2016 [1].

Such a high penetration of solar generation increases concerns in supply and demand balancing given that solar generation is highly dependent on weather conditions and cloud shading. To address solar generation uncertainties, stationary and mobile storages can be connected to the solar generation as buffers [2], [3]. Even though Battery Energy Storage System (BESS) and gridable Electrical Vehicles (EVs) can compensate for the solar generation fluctuation, a better prediction of the solar generation helps to reduce the size of the BESS and the number of EVs. Furthermore, an accurate prediction of the solar generation helps to improve the operational cost of Demand Side Management (DSM) [4], [5].

The prediction of solar power generation is generally classified into two categories, i.e., model-based prediction and model-free prediction. In model-based predictions, factors that impact solar generation are the parameters of the famous I-V curve used for the prediction of the solar power [6]. Huang et al. studied the short term solar irradiance change based on cloud motion image processing [7]. Capizzi et al. investigated environmental parameters such as humidity and temperature to achieve a better estimation based on the correlation of the parameters with solar generation [8]. These methods give precise predictions of solar generation relying on the extensive knowledge of the environmental conditions. It incurs additional expense in installing sensing and communication facilities, which is feasible for utility solar farms with aggregated solar generation at MW level. However, in distributed solar generation of smaller sizes, it is common that the solar panels do not include temperature, irradiation and humidity sensors.

The majority of solar power prediction methods rely on model-free computational methods. Solar power generation data is normally stored for research and the predictions are made based on this historical data, with time series analysis and machine learning techniques. Huang et al. made prediction of the solar irradiance based on Auto-Regressive Moving Average (ARMA) model [9]. Ruffing et al. made multi-step solar prediction with Echo State Network (ESN) in [10]. ARMA and ESN only predict for a small time horizon which cannot satisfy applications requiring day-ahead solar prediction. Negash et al. studied the solar prediction with NARX model [11]. Deng et al. estimated solar generation with Support Vector Machine (SVM) [12]. The performance of these methods are highly dependent on long-term data collection. The longer the data collection period, the more likely the future solar generation may fall into a historical pattern. As these methods do not need additional sensors, the prediction accuracy can be limited if the weather changes abruptly or if the long-term data collection is not available.

This paper studies a typical scenario of roof-top solar power generation at residential homes and commercial buildings where the real-time temperature, irradiation and humidity information is not available. We propose a two-tier method for solar power generation prediction using historical power data alone and validated it for 35kW capacity solar generation in the UCLA Microgrid. To study how different global-tier predictions may affect the overall performance of the two-tier prediction, global-tier day-ahead predictions are performed with two heuristic methods, i.e., weighted k-Nearest

This work has been sponsored in part by a grant from the LADWP/DOE fund 20699 & 20686, (Smart Grid Regional Demonstration Project).

Neighbors (k-NN) and Neural Network (NN). In local-tier, adaptive real-time corrections based on residual analysis is applied to improve the day-ahead prediction results. The contribution of the paper is three-fold: first, compared with traditional day-ahead predictions, the proposed method achieves higher accuracy while maintaining the same prediction length. Second, unlike most real-time models, the proposed local-tier prediction is not heuristic. The analytical local-tier has a clear physical meaning and it has a low computation cost. The local-tier can be combined with other day-ahead prediction methods to improve the prediction accuracy. Finally, reasons for why NN-based two-tier prediction method outperforms weighted k-NN based counterpart are discussed, which sheds some light on the principles to design the global-tier method.

The remainder of this paper is organized as follows: Section II articulates global day-ahead prediction methods. Both weighted k-NN and NN model are used for predictions. It is followed by presenting the local real-time residual analysis based correction method in Section III. Section IV discusses the experimental setup and result analysis. Finally, conclusions are drawn in Section V.

## II. GLOBAL-TIER DAY-AHEAD PREDICTION

The proposed prediction method has two tiers, i.e., a global-tier for day-ahead prediction and a local-tier for real-time correction. We assume that the solar power data is sampled with a sampling interval of $T_s$ for $N$ days, we have a sequence of measured solar power data $P$:

$$P = \{P_1(0), P_1(T_s), \ldots, P_i(mT_s), \ldots, P_N(MT_s)\} \quad (1)$$

where $m \in \{0,1,\ldots,M\}$, $i \in \{1,2,\ldots,N\}$, $P_i(mT_s)$ denotes the measured power of day $i$ at time instant $mT_s$, and $M$ is maximum samples per day.

The two popular heuristic methods that are investigated, weighted k-NN and NN use models that need to be trained before prediction. For both models we assume that collected data is arranged in a chronical order and the first $rN$ days of data is used for the training purpose with $r \in (0,1)$. The $(1-r)N$ days of the remaining data is used to validate the models. We use $x_t$ and $y_t$ as the training input and output sets for the models, and $x_v$ and $y_v$ as the verification input and verification output sets. Then we have:

$$x_t, y_t \in \{P_1(0), P_1(T_s), \ldots, P_1(MT_s), P_2(0), \ldots, P_{rN}(MT_s)\} \quad (2)$$

We assume that $\hat{P}_g(mT_s) \in y_v, g > rN$ where $\hat{P}_g(mT_s)$ is the predicted solar generation power in day $g$ at time $mT_s$, then we have:

$$x_v \in \{P_i(mT_s) | 0 < i < g, \ 0 \leq m \leq M\} \quad (3)$$

The main idea behind the studied weighted k-NN and NN approach is that the future data is a subsequence of the historical data, i.e.,

$$x_t^p \in x_t, \ y_t^p \in y_t, \ x_v^q \in x_v, \ y_v^q \in y_v \quad (4)$$

where $x_t^p$ and $x_v^q$, $y_t^p$ and $y_v^q$ are respectively $p^{th}$ and $q^{th}$ elements of the sets. The day-ahead prediction method problem is to predict the solar generation for the next day based on the historical data, and it is formulated as follows:

$$\hat{y}_t^p = f(x_t^p), \ \hat{y}_v^q = f(x_v^q) \quad (5)$$

where $f$ denotes the mapping from historical values to predicted values and $\hat{y}_t^p$, $\hat{y}_v^q$ are the predictions of $y_t^p$ and $y_v^q$. The problem is subsequently formulated as how to map the historical data to day-ahead predictions.

To evaluate the performance of the prediction, we use the Root Mean Squared Error (RMSE) defined as follows:

$$RMSE = \sqrt{\frac{1}{M}\sum_{m=0}^{M}(\hat{P}_g(mT_s) - P_g(mT_s))^2} \quad (6)$$

Next the weighted k-NN and NN are studied for global-tier day-ahead prediction. The study is used to demonstrate that the proposed local-tier adaptive prediction method works well with general global-tier prediction methods and to show how different global-tier predictions affect the performance of two-tier prediction.

### A. Weighted k-NN model

Weighted k-NN is an evolution of the machine learning algorithm k-NN [13]. The idea of weighted k-NN is to search for the k-nearest (most similar) patterns and combine them with higher weighting to more similar ones for prediction.

The weighted k-NN algorithm is tabulated in Algorithm 1, the data for training the weighted k-NN model is:

$$\begin{aligned} x_t^i &= \{P_{i-D}(0), \ldots, P_{i-D}(MT_s), P_{i-D+1}(0), \ldots, P_{i-1}(MT_s)\} \\ y_t^i &= \{P_i(0), P_i(T_s), \ldots, P_i(MT_s)\} \end{aligned} \quad (7)$$

Note that the idea behind the algorithm is that it relates the specific day's prediction with the observed data for the previous $D$ days.

---

**Algorithm 1:** Day-ahead Prediction with Weighted $k-NN$

**Data:** $x_t, y_t$
**Input:** $x_v$
**Output:** $\hat{y}_v$

for each $x_v^j \in x_v$ do
  for each $x_t^i \in x_t$ do
  $\quad dis(i,j) \leftarrow \|x_v^j - x_t^i\|$;
  end
  $idx(l) \leftarrow i$, where $dis(i,j)$ is the $l^{th}$ smallest element;
  for each $l \in [1, 2, .., k]$ do
  $\quad w(l) \leftarrow \frac{dis(idx(k+1),j) - dis(idx(l),j)}{dis(idx(k+1),j) - dis(idx(1),j)}$
  end
  $\hat{y}_v^j \leftarrow \frac{1}{\sum_{l=1}^{k} w(l)} \sum_{l=1}^{k} w(l) y_t^{idx(l)}$;
end

---

It also needs to be pointed out that the distance we are using in the weighted k-NN is the Euclidian distance. It reveals the similarity between data sets. Furthermore, the weighted k-NN algorithm differs from traditional k-NN in a way that it gives more weight to more similar patterns. Typically, weighted k-NN gives better prediction results compared to k-NN.

## B. NN Model

NN is an effective model for predictions and pattern recognitions. The essential idea behind the model is to use multi-layer neural networks in capturing the high dimensional nonlinear mapping between inputs and outputs. The structure of the NN used in this prediction is shown in Fig. 1.

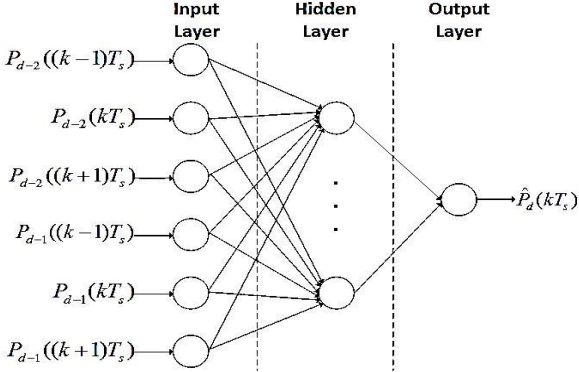

Fig. 1. Topological structure of the NN

As shown in Fig. 1, the output power at time instance $kT_s$ of day $d$ is related to the previous two days. The NN model uses three layers with a hidden layer in between the input and output layers. For hidden layer neuron number, there is a large literature and one of the data driven methods is documented in [14]. We will discuss the hidden layer neuron number in Section IV.

## III. LOCAL-TIER REAL-TIME CORRECTION

The global-tier day-ahead solar power prediction is made using one of the two methods discussed in the previous section. However, their accuracy must be limited as we are making a day-ahead prediction without the knowledge of the solar generation on the predicted day. Using the real-time solar generation, prediction results can be substantially improved. Assume that at time instance $mT_s$ of day $i$, we have the measured data $P_i(mT_s)$ and the day-ahead predicted data $\hat{P}_i(mT_s)$. The residual $R(m)$ is defined as:

$$R(m) = \hat{P}_i(mT_s) - P_i(mT_s) \qquad (8)$$

The index in $R(m)$ omitted as we are making real-time correction.

The underlining idea of local real-time correction is to extract the low frequency components in residual sequence for compensation. See [15] for a detailed discussion on residual analysis. We define the sequence of the residuals to be

$$S_n^m = \{R(m-n+1), R(m-n+2), \ldots, R(m)\} \qquad (9)$$

where $m \geq n-1$ and $S_n^m$ is the residual sequence at time instant $mT_s$ with length $n$. We use Discrete Fourier Series (DFS) to extract the low frequency component of the residual sequence. The DFS of the $S_n^m$ is represented as:

$$S_n^m(k) = \sum_{i=0}^{L} \left( a_i \cos(k\frac{2\pi i}{T_s}) + b_i \sin(k\frac{2\pi i}{T_s}) \right) \qquad (10)$$

where $k = [1,2,\ldots,n]$; $S_n^m(k)$ is the $k^{th}$ component of $S_n^m$; $a_i$ and $b_i$ are the DFS coefficients representing the $i^{th}$ frequency components. The sampling time interval $T_s$ determines the bandwidth of signal and the sampling point $n$ determines the resolution of the discrete frequency components [16]. $L$ represents the maximum allowable frequency component. The above defined terms are represented in compact matrix form as follows:

$$S = [S_n^m(1) \ S_n^m(2) \ldots S_n^m(n)]^T \qquad (11)$$

$$C = [a_0 \ b_0 \ a_1 \ldots b_L]^T \qquad (12)$$

$$F_{uv} = \begin{cases} \cos\left(v\frac{2\pi(\overline{u/2}-1)}{T_s}\right), & mod(u,2) = 0 \\ \sin\left(v\frac{2\pi(\overline{u/2}-1)}{T_s}\right), & mod(u,2) = 1 \end{cases} \qquad (13)$$

where $u = [1,2,\ldots,2L]$, $v = [1,2,\ldots,n]$. The coefficients can be estimated using the Least-Square method as follows:

$$C = (F^T F)^{-1} F^T S \qquad (14)$$

Given an estimation of the coefficient matrix, we can compute $S_n^m(k)$ according to (10) and use it for the correction of the future day-ahead predictions. Using the computed low frequency components in (10), we dynamically update the day-ahead prediction as follows:

$$\hat{P}_i^c(lT_s) = \hat{P}_i(lT_s) + S_n^m(mod((l-k),k)+1) \qquad (15)$$

where $l = [m+1, m+2, \ldots, M]$ and $\hat{P}_i^c(lT_s)$ is the corrected prediction for day $i$ at time $lT_s$ and $\hat{P}_i(lT_s)$ is the global-tier estimation result. The local-tier correction takes into account the real-time trend and adaptively updates the day-ahead estimated value. In the following sections we will see how lower frequency components of the residual sequence contribute to more accurate solar power predictions.

## IV. EXPERIMENTAL SETUP AND RESULTS ANALYSIS

In this section, UCLA Microgrid is used as the testbed for an evaluation of the proposed algorithms. Solar generation data is collected from a 35kW capacity solar installation. For further information about solar generation at UCLA, online access can be found [17]. Fifty days of solar data was collected from Feb. 15th for training and verification of the algorithm. 60 percent of the data is used for training purposes, 20 percent for model tuning, and the remaining 20 percent is used for testing and analysis. Data is collected with $T_s = 15$ min.

### A. Model Tuning

For weighted k-NN day-ahead prediction, parameters $D$ and $k$ need to be tuned and determined. The comparison of the prediction accuracy for different $D$ and $k$ is carried out with 20 percent of collected data. Table I shows how the average RMSE varies with $D$. Test are done for $D$ varying from 1 to 8 with each RMSE representing the daily average RMSE of the verification data. The test for varying $k$ has also been carried out and tested from 2 to 4 (if $k = 1$ then weighted k-NN decays to k-NN) with $k = 2$ the smallest. Comparison results show that the weighted k-NN day-ahead prediction works best with $k = 2$ and $D = 5$. $k = 2$ shows that best result is generated by combining only the two most similar days, which indicates that solar generations actually falls into very different patterns.

For NN based day-ahead prediction, the number of hidden layer neurons needs to be determined. Same 20 percent of collected data is used for determination of the hidden layer neuron number. However, as is well-known that training results for NN are determined by the random initial value, and sometimes bad initial values may drag the NN into local optimum. Taking the randomness into account, we train the NN 10 times with Levenberg-Marquardt method and get the average value of RMSE. From the results shown in Table II, we found that NN with 6 hidden layer neurons gives the best prediction results. When the hidden layer neurons are greater than 6, there is no significant improvement in accuracy due to the problem of over-fitting.

TABLE I. COMPARIONS OF RMSE OVER $D$

| $D$ | 1 | 2 | 3 | 4 |
|---|---|---|---|---|
| $RMSE^a$ | 0.669 | 0.572 | 0.649 | 0.420 |
| $D$ | 5 | 6 | 7 | 8 |
| $RMSE$ | 0.405 | 1 | 0.939 | 0.865 |

a. RMSE 4943.6 is normalized to 1

TABLE II. COMPARISON OF RMSE OVER HIDDEN LAYER NEURONS

| $N$ | 3 | 4 | 5 |
|---|---|---|---|
| $RMSE^a$ | 1 | 0.959 | 0.943 |
| $N$ | 6 | 7 | 8 |
| $RMSE$ | 0.907 | 0.940 | 0.956 |

a. RMSE 2499.5 is normalized to 1

We have determined the optimal parameters for the day-ahead predictions using numerical and not analytical methods. It implies that the optimal parameters are perhaps data-dependent. Therefore, with longer period of data collection, we may finally reach to a set of stable data-driven optimal parameters for day-ahead predictions. On local-tier real-time correction side, the sequence length $n$ is chosen to be 8, i.e., a two-hour observation. The length $n$ determines the resolution in its frequency spectrum. Parameter $L$ is selected as 2, which only accepts the first two frequency component of the residual sequence. Increasing $L$ means adding additional higher frequency variation into the correction signal. Note that $L$ cannot be larger than $n/2$ otherwise the Least Squares problem becomes overdetermined.

### B. Prediction Results Analysis

Fig. 2 shows the overall prediction accuracy for all test days. The four curves indicate the RMSE of predicting using weighted k-NN alone, NN alone, two-tier prediction with weighted k-NN as the global-tier prediction and two-tier prediction with NN as the global-tier prediction respectively. The averaged RMSE over test days of each curve is also plotted. It is observed from the results that weighted k-NN and NN based day-ahead prediction gives similar prediction accuracy. However, when combined with the local-tier correction, weighted k-NN and NN based two-tier predictions show improvements of 28.02 percent and 40.36 percent respectively compared with using the day-ahead prediction alone.

Besides the overall performance of the prediction, it is desirable to have a closer look on how the prediction methods work for one particular test day. Fig. 3 shows the comparison results of different prediction methods on a sunny day. The results are separated into two figures, each of which makes the comparison using day-ahead prediction only with its corresponding two-tier prediction. It is observed that the day-ahead prediction of both weighted k-NN and NN aligns with the real power measurement pretty well. However, two-tier prediction further approaches the real measurement. Improvements of 30.34 percent and 51.96 percent are made respectively with the two-tier prediction. Note that as the sampling frequency is low and we are filtering out the high frequency component within the bandwidth of half of the sampling frequency, this causes some observable overshoots are in the two-tier prediction method. Theoretically, the prediction results can be further improved by increasing the sampling frequency.

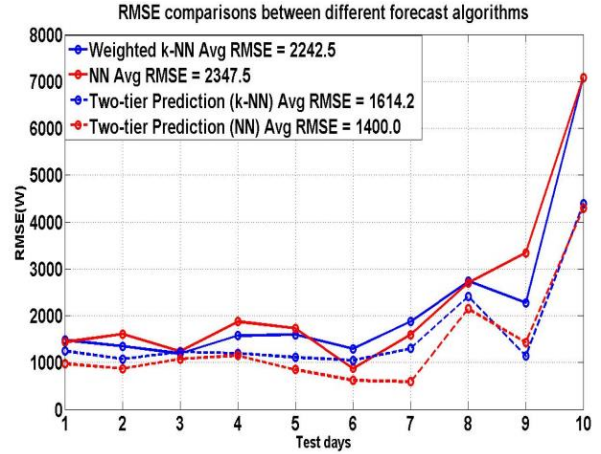

Fig. 2. RMSE comparison between different forecast algorithms

Apart from the observation of prediction results on a sunny day, it is also interesting to see how different prediction methods work on a cloudy day. Fig. 4 shows the comparison for a cloudy day. The day-ahead prediction of weighted k-NN and NN are poor. However, with the proposed two-tier prediction method, the predicted value actually approaches the real measurements well. It picks up the low frequency residual of the real measurements and day-ahead predictions. Note that there still exists undershoot in the prediction partly resultingfrom the limited bandwidth. The overshoot part below zero is set to zero in the two-tier prediction methods, so there is a sharp drop in two-tier prediction curve when it is close to zero. Fig. 4 shows that the two-tier predictions achieves 37.87 percent and 39.37 percent relative improvements and 2679.2 and 2788.7 absolute improvements on RMSE over the day-ahead prediction alone.

### C. Comparisons of weighted k-NN and NN based Two-Tier Predictions

It is observed that the local-tier prediction works with both the global-tier prediction methods. Using global-tier day-ahead prediction alone, the weighted k-NN and NN have a similar prediction accuracy. However, NN based two-tier prediction generally outperforms its weighted k-NN counterpart for test-day average and specific sunny/cloudy day. The difference between the two may result from the nature of the two algorithms: the NN nonlinearly approximates solar generation curve while weighted k-NN is a linear combination of historical data. Though they generates same level of error in the global-tier, NN introduces less high frequency error compared with weighted k-NN. Given that the local-tier correction filters out high frequency error and is essentially a low frequency compensator, this explains why NN based two-tier prediction have a better performance.

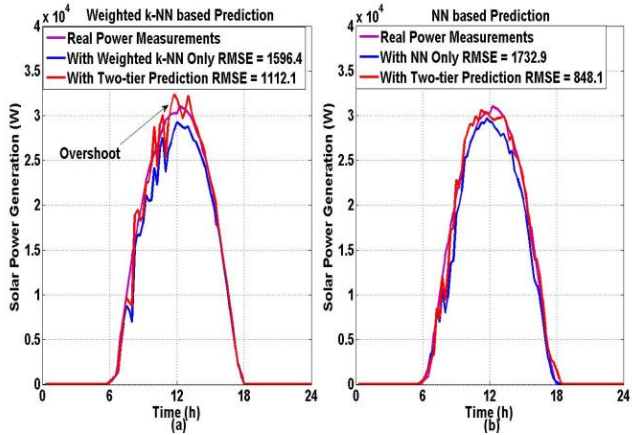

Fig. 3. RMSE comparisons in a sunny day. (a) comparions with the day-ahead predictions using weighted k-NN, (b) comparisons with day-ahead predictions using NN

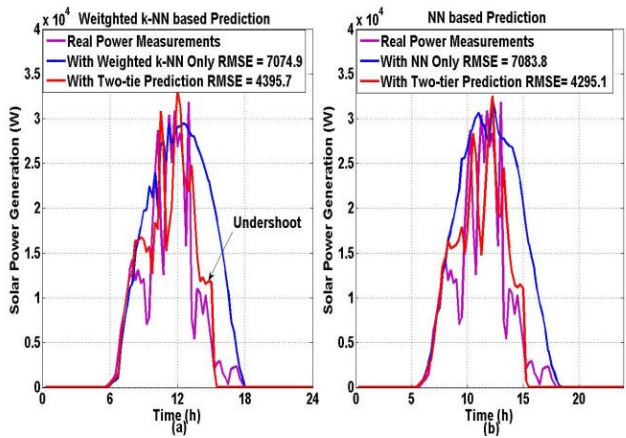

Fig. 4. RMSE comparisons in a cloudy day. (a) comparions with the day-ahead predictions using weighted k-NN, (b) comparisons with day-ahead predictions using NN

## V. CONCLUSIONS

In this paper, the problem of solar power prediction with limited sensing resource is described and analyzed. We propose a two-tier prediction that combines the global-tier day-ahead predictions with local-tier real-time residual analysis. Experimental results show that weighted k-NN based and NN based two-tier methods achieve 28 percent and 40 percent accuracy improvements respectively compared with their day-ahead counterparts. Furthermore, case studies in a typical sunny and a cloudy day are carried out, which shows that the proposed method is particularly effective on days when solar generation is variable. Finally, it is also observed that though weighted k-NN and NN achieve similar accuracy in the global-tier, NN based two-tier prediction generally outperforms its weighted k-NN based counterpart. Comparisons and analysis of the differences shed light on global-tier algorithm design. The local-tier method can be combined with other global-tier methods for achieving higher prediction accuracy. Thus, the proposed two-tier method provides good basis for smart grid applications that require long-term solar predictions.